\documentclass{article}

\title{A Solution Technique for Quantum Mechanical Differential Equations Using Multiple Complex Planes.}
\author{Robert Ducharme}

\begin{document}

\maketitle

\centerline{151 Fairhills Dr., Ypsilanti, MI 48197}
\centerline{E-mail: ducharme01@comcast.net}

\begin{abstract} 
It is shown fields that cannot be represented over one complex plane can be further decomposed for representation over multiple complex planes. This finding is demonstrated here by solving of the Schr\"{o}dinger equation for the hydrogen atom in a complex space containing three complex planes. The complex coordinate system is generated from real coordinates using an isometric transformation. One plane is applied to mix energy and time; the other two planes are used to represent the z-component of angular momentum of the electron. The eigensolutions of the Schr\"{o}dinger equation are shown to be holomorphic in the complex planes. 
\end{abstract}

\section{Introduction}
Most practical applications of conformal mapping \cite{ZN} have been limited to one complex plane. Liouville's theorem \cite{DEB} in fact shows that higher dimensional conformal maps are possible but must be composed of translations, similarities, orthogonal transformations and inversions. The purpose of this paper is to show problems that cannot be represented over one complex plane can generally be decomposed for representation over multiple complex planes. Candidates for decomposition in to multiple complex planes may include problems that have more than two dimensions in real space or two-dimensional problems that fail to satisfy the Cauchy-Riemann equations. The conformal transformations to be presented in this paper are isometric and therefore fall under Liouville's theorem. 

An hydrogen atom consists of a single electron with a spatial displacement $x_i$ $(i=1,2,3)$ from the center of mass of the atom. The source of the Coulomb field binding the electron to the atom is the much heavier proton. Both the electron and the proton may be assumed to exist at the same time t. The objective of section 2 of this paper, taking the Schr\"{o}dinger equation of the hydrogen atom as an illustrative example \cite{RJD1}, is to introduce an isometric conformal mapping from a real $(x_i,t)$ coordinate system to a complex $(z_i,\tau)$ coordinate system. The inverse, differential and inverse differential forms of the transformation are also presented. 

The $(z_i,\tau)$ coordinate system contains three complex planes $z_1$, $z_2$ and $\tau$. The $z_3$ coordinate is pure real. In order for complex derivatives of a continuously differentiable function $f(z_i,\tau)$  to be holomorphic, the value of the derivative cannot depend on the direction of differentiation in each complex plane. The set of Cauchy-Riemann equations expressing this condition is determined in section 3. 

One of the interesting features of the $(z_i,\tau)$ coordinate system is that the $x_1x_2$-plane is decomposed into two complex planes $z_1$ and $z_2$. This means that each $(x_1,x_2)$ point in the real plane has one image point in the $z_1$ plane and a second in the $z_2$ plane. The reason for this is that owing to the constraint of the Cauchy-Riemann equations not every function $f(x_1,x_2)$ can be usefully represented in a single complex plane. It will be shown in section 3 that the transform $f(z_1,z_2)$ of any function $f(x_1,x_2)$ will be holomorphic in both the $z_1$-plane with $z_2$ held constant and the $z_2$-plane with $z_1$ held constant. This further implies that any function of $N$ real coordinates will have an holomorphic representation in complex space consisting of $N$ complex planes providing both spaces are related through an isometric transformation. 

The hydrogen atom is of interest in this paper for two reasons. One is that Schr\"{o}dinger equation generates closed form analytical solutions for the hydrogen atom in spherical polar coordinates. The other is that the hydrogenic wavefunctions can have angular momentum.  By contrast, it is known that a field over a one complex plane must be irrotational. It is therefore a test of the aforementioned decomposition principle to see if the problem of the hydrogen atom can be represented across multiple complex planes.

In section 4, the Schr\"{o}dinger equation of the hydrogen atom is transformed into both complex Cartesian and complex spherical polar coordinates. One advantage of complex versus real spherical polar coordinates is that there are no trigonometric functions in them leading to a simpler Laplacian term in the  Schr\"{o}dinger equation. It is shown the use of complex spherical polar coordinates to eliminate trigonometric functions alongside an application of the complex time coordinate to remove the exponential term in the radial component of the hydrogenic wavefunctions leads to a much simplified solution process in complex space compared to real space. It is further shown that any of the results obtained in complex space are readily transformed back into real space. In particular, the hydrogenic wavefunctions obtained in complex space are found to transform into the generally accepted results \cite{DFL} in real spherical polar coordinates.  

\section{Complex Space}
The task ahead is to present an isometric conformal transformation relating a real $(x_i,t)$ coordinate system $(i = 1,2,3)$ and a complex $(z_i,\tau)$ coordinate system. This mapping is to be applied in section 4 to a hydrogen atom consisting of an electron of total energy E at a separation $r (=|x_i|)$ from the proton. It can be expressed in the form
\begin{equation} \label{eq: conftrans1}
z_1 = \frac{x_1+\imath x_2}{\sqrt{2}}, \quad z_2 = \frac{x_2+\imath x_1}{\sqrt{2}}  
\end{equation}
\begin{equation} \label{eq: conftrans2}
z_3 = x_3, \quad \tau = t - \imath \frac{\hbar |x_i|}{\alpha E}
\end{equation}
where $\alpha$ is a constant and $\hbar$ is Planck's constant divided by $2\pi$. It will further be assumed that the components of the $(z_i,\tau)$ coordinate system are independent such that
\begin{equation} \label{eq: metric}
\frac{\partial z_i}{\partial z_j} = \delta_{ij}, \quad \frac{\partial \tau}{\partial z_i} = 0 
\end{equation}
where $\delta_{ij}$ is the Kronecker delta. It is important to be clear that every point in the $(x_1, x_2)$ plane has an image in both the $z_1$ and $z_2$ complex planes. Eq. (\ref{eq: conftrans1}) therefore constitutes a decomposition of a real plane into two complex planes.

In the $(x_i,t)$-coordinate system, the electron has a spatial displacement $x_i$ from the proton but shares the same world time t. Similarly, in the $(z_i,\tau)$-coordinate system, the electron has a displacement $z_i = x_i$ from the proton but shares the same complex time $\tau$. It can be shown using eqs. (\ref{eq: conftrans1}) and (\ref{eq: conftrans2}) that
\begin{equation} \label{eq: isometric}
z_1^* z_1 + z_2^* z_2 + z_3^*z_3 = x_1^2 + x_2^2 + x_3^2
\end{equation}
\begin{equation} \label{eq: correlated}
\frac{z_1}{z_2^*} = \frac{z_2}{z_1^*} = \imath
\end{equation}
\begin{equation} \label{eq: modz1}
-2 \imath z_1 z_2 = x_1^2 + x_2^2
\end{equation}
where $z_i^*$ denotes the complex conjugate of $z_i$. Eq. (\ref{eq: isometric}) alongside the understanding that the electron and proton share the same value of $\tau$ confirms the isometric nature of the transformation. 

The inverses of the transformation equations (\ref{eq: conftrans1}) can be written
\begin{equation} \label{eq: inv_ct}
x_1 = \frac{z_1 - \imath z_2}{\sqrt{2}}, \quad x_2 = \frac{z_2 - \imath z_1}{\sqrt{2}}
\end{equation}
\begin{equation} \label{eq: inv_ict1}
x_{i} = z_{i}, \quad t = \tau + \imath \frac{\hbar}{E}\frac{|z_i|}{\alpha} 
\end{equation}
where $|z_i| = |x_i|$ 

Derivatives with respect to $(z_i,\tau)$ coordinates can be evaluated using the chain rule for partial differentiation to give
\begin{equation} \label{eq: complexDiff1}
\frac{\partial}{\partial z_1} 
= \frac{1}{\sqrt{2}} \left[ \frac{\partial}{\partial x_1} - \imath \frac{\partial}{\partial x_2} + \frac{(x_1-\imath x_2)}{\alpha |x_i|}   \frac{\imath \hbar}{E}  \frac{\partial}{\partial t} \right],
\end{equation}
\begin{equation} \label{eq: complexDiff2}
\frac{\partial}{\partial z_2} 
= \frac{1}{\sqrt{2}} \left[ \frac{\partial}{\partial x_2} - \imath \frac{\partial}{\partial x_1} + \frac{(x_2-\imath x_1)}{\alpha |x_i|}   \frac{\imath \hbar}{E}  \frac{\partial}{\partial t} \right],
\end{equation}
\begin{equation} \label{eq: complexDiff3}
\frac{\partial}{\partial z_3} 
= \frac{\partial}{\partial x_3} + \frac{x_3}{\alpha |x_i|}   \frac{\imath \hbar}{E}  \frac{\partial}{\partial t}, \quad \frac{\partial}{\partial \tau} = \frac{\partial}{\partial t}
\end{equation}
having used eqs. (\ref{eq: metric}) and (\ref{eq: inv_ct}). Eq. (\ref{eq: conftrans1}) can be used alongside these results to give
\begin{equation} \label{eq: complexProd}
z_i\frac{\partial}{\partial z_i} 
= x_i\frac{\partial}{\partial x_i} + \frac{|x_i|}{\alpha}\frac{\imath \hbar}{E}  \frac{\partial}{\partial t}
\end{equation}
These result will prove useful later.

The inverses of eqs. (\ref{eq: complexDiff1}) and (\ref{eq: complexDiff2}) are
\begin{equation} \label{eq: invComplexDiff1}
\frac{\partial}{\partial x_1} 
= \frac{1}{\sqrt{2}} \left[ \frac{\partial}{\partial z_1} + \imath \frac{\partial}{\partial z_2} - \frac{(z_1-\imath z_2)}{\alpha |z_i|}   \frac{\imath \hbar}{E}  \frac{\partial}{\partial \tau} \right],
\end{equation}
\begin{equation} \label{eq: invComplexDiff2}
\frac{\partial}{\partial x_2} 
= \frac{1}{\sqrt{2}} \left[ \frac{\partial}{\partial z_2} + \imath \frac{\partial}{\partial z_1} - \frac{(z_2-\imath z_1)}{\alpha |z_i|}   \frac{\imath \hbar}{E}  \frac{\partial}{\partial \tau} \right],
\end{equation}
\begin{equation} \label{eq: invComplexDiff3}
\frac{\partial}{\partial x_3} 
= \frac{\partial}{\partial z_3} - \frac{z_3}{\alpha |z_i|}   \frac{\imath \hbar}{E}  \frac{\partial}{\partial \tau}, \quad \frac{\partial}{\partial t} = \frac{\partial}{\partial \tau}
\end{equation}
Eqs. (\ref{eq: complexDiff1}) through (\ref{eq: complexDiff3}) can be used to validate eq. (\ref{eq: metric}) confirming the $(z_i,\tau)$ coordinates are all independent of each other. They also give
\begin{equation} \label{eq: complexDiff5}
\frac{\partial z_i}{\partial z_i^*}=\frac{\partial \tau}{\partial \tau^*} = 0
\end{equation}
showing complex conjugate coordinates are independent of each other. 

It is important to be mindful of the difference between the complex conjugate of a derivative and derivatives with respect to complex conjugate variables. For example, the complex conjugate of eq. (\ref{eq: complexDiff1}) is
\begin{equation} \label{eq: complexDiff1_conj}
\frac{\partial^*}{\partial z_1} 
= \frac{1}{\sqrt{2}} \left[ \frac{\partial}{\partial x_1} + \imath \frac{\partial}{\partial x_2} + \frac{(x_1+\imath x_2)}{\alpha |x_i|}   \frac{\imath \hbar}{E}  \frac{\partial}{\partial t} \right],
\end{equation}
By contrast, it can be shown using the chain rule of partial differentiation that
\begin{equation} \label{eq: complexDiff1_zconj}
\frac{\partial}{\partial z_1^*} 
= \frac{1}{\sqrt{2}} \left[ \frac{\partial}{\partial x_1} + \imath \frac{\partial}{\partial x_2} - \frac{(x_1+\imath x_2)}{\alpha |x_i|}   \frac{\imath \hbar}{E}  \frac{\partial}{\partial t} \right],
\end{equation}
Note the sign difference between these two results. The conjugate derivatives used in this paper will all be of the form $\partial^* / \partial z_i$. Conjugate derivatives of the form $\partial / \partial z_i^*$ have been used in previous work both to represent ladder operators in the treatment of harmonic oscillators \cite{RJD2} and to demonstrate the formal elimination of potential terms from quantum mechanical equations \cite{RJD3} (compare to \cite{FF}).

\section{The Cauchy-Riemann Equations}
The purpose of this section is to follow through the consequences of the Cauchy-Riemann equations for a continuously differentiable function $f(x_i, t)$ transformed into each of the complex planes $z_1$, $z_2$ and $\tau$. These conditions are can be expressed as
\begin{equation} \label{eq: holo1}
\left(\frac{\partial f}{\partial t} + \imath \frac{\partial f}{\partial s} \right)_{z_1, z_2, z_3} = \left(\frac{\partial f}{\partial \tau^*} \right)_{z_1, z_2, z_3} = 0, 
\end{equation}
\begin{equation} \label{eq: holo2}
\left(\frac{\partial f}{\partial x_1} + \imath \frac{\partial f}{\partial x_2} \right)_{\tau, z_2, z_3} = \left(\frac{\partial f}{\partial z_1^*} \right)_{\tau, z_2, z_3} = 0, 
\end{equation}
\begin{equation} \label{eq: holo3}
\left(\frac{\partial f}{\partial x_2} + \imath \frac{\partial f}{\partial x_1} \right)_{\tau, z_1, z_3} = \left(\frac{\partial f}{\partial z_2^*} \right)_{\tau, z_1, z_3} = 0, 
\end{equation}
having set
\begin{equation} \label{eq: sdef}
s = - \frac{\hbar r}{\alpha E} 
\end{equation}
such that $\tau = t + \imath s$. Here, the complex variables outside the brackets must be held constant during the differentiation. 

Eqs. (\ref{eq: holo1}) can also be written
\begin{equation} \label{eq: holo1_alternative}
\left(\frac{\partial^2 f}{\partial t^2} + \frac{\alpha^2 E^2}{\hbar^2} \frac{\partial^2 f}{\partial r^2} = 0 \right)_{z_1, z_2, z_3} = 0, 
\end{equation}
It simplifies slightly if $f$ can be assumed to take the product form
\begin{equation} \label{eq: f_productSolution}
f(z_i,\tau) = f_z(z_i)f_{\tau}(\tau)
\end{equation}
to give 
\begin{equation} \label{eq: holo1_simple}
\frac{\partial^2 f_{\tau}}{\partial t^2} + \frac{\alpha^2 E^2}{\hbar^2} \frac{\partial^2 f_{\tau}}{\partial r^2} = 0
\end{equation}
though eq. ($\ref{eq: holo1_alternative}$) can also be evaluated directly and a product form is not a requirement. Any function $f$ will satisfy eq. (\ref{eq: holo2}) in the form
\begin{equation} \label{eq: holo2_simple}
\left(\frac{\partial f}{\partial z_1^*} \right)_{z_2, z_3, \tau} = 0
\end{equation}
since $z_1$ and $z_1^*$ are independent and all of $z_2$, $z_3$ and $\tau$ are held constant. Similarly, any function $f$ will satisfy eq. (\ref{eq: holo3}) in the form
\begin{equation} \label{eq: holo3_simple}
\left(\frac{\partial f}{\partial z_2^*} \right)_{z_1, z_3, \tau} = 0, 
\end{equation}
since $z_2$ and $z_2^*$ are independent and all of $z_1$, $z_3$ and $\tau$ are held constant. It is concluded on this basis that any continuously differentiable function $f$ transformed from $(x_i,t)$-space to $(z_i,\tau)$-space will be holomorphic in the $(z_i,\tau)$-space providing it satisfies eq. (\ref{eq: holo1_alternative}).

Derivatives with respect to real $(x_i,t)$-coordinates that are constrained to complex planes can be evaluated using eqs. (\ref{eq: invComplexDiff1}) through (\ref{eq: invComplexDiff3}) to give
\begin{equation} \label{eq: constrained_deriv1}
\left(\frac{\partial^{m+n} f}{\partial x_1^m \partial x_2^n} \right)_{z_2,z_3,\tau} = \imath^n \left(\frac{1}{\sqrt{2}}\frac{\partial }{\partial z_1} \right)^{m+n} f, 
\end{equation}
\begin{equation} \label{eq: constrained_deriv2}
\left(\frac{\partial^{m+n} f}{\partial x_1^m \partial x_2^n} \right)_{z_1,z_3,\tau} = \imath^m \left( \frac{1}{\sqrt{2}}\frac{\partial }{\partial z_2} \right)^{m+n} f, 
\end{equation}
\begin{equation} \label{eq: constrained_deriv3}
\left(\frac{\partial f}{\partial x_3} \right)_{z_1,z_2,\tau} =  \frac{\partial f}{\partial z_3}, \quad \left(\frac{\partial f}{\partial t} \right)_{z_1,z_2,z_3} =  \frac{\partial f}{\partial \tau}
\end{equation}
where $m, n = 0, 1, 2 ...$. Eqs. (\ref{eq: constrained_deriv1}) and (\ref{eq: constrained_deriv2}) show that any function $f_z(x_i)$ will satisfy Laplace's equation
\begin{equation} \label{eq: laplace}
\left(\frac{\partial^2 f}{\partial x_1^2} + \frac{\partial^2 f}{\partial x_2^2} \right)_{z_3, \tau} =  0
\end{equation}
in both  the $z_1$ and $z_2$ complex planes irrespective of the form of $f$. This is, in fact, just an alternative statement of the Cauchy-Riemann equations (\ref{eq: holo2_simple}) and (\ref{eq: holo3_simple}).

\section{The Hydrogen Atom}
The Schr\"{o}dinger equation determining the wavefunction $\Psi(x_i, t)$ for an electron of mass $m_e$ bound in the Coulomb field of a proton can be expressed in the form
\begin{equation} \label{eq: schrod1}
-\frac{\hbar^2}{2 m_e} \nabla^2 \Psi - \frac{e^2}{4 \pi \epsilon_0 r}\Psi = E\Psi
\end{equation}
In this, $\nabla^2 ( = \partial^2 / \partial x_i^2)$ is the Laplacian operator, -e is the charge on the electron, $\epsilon_0$ is the permittivity of free space and
\begin{equation} \label{eq: schrod2}
E\Psi = \imath \hbar \frac{\partial \Psi}{\partial t} 
\end{equation}
gives the total energy of the electron.

In developing the connection between complex $(z_i,\tau)$-coordinates and the hydrogen atom, eqs. (\ref{eq: complexDiff1}) through (\ref{eq: complexDiff3}) and (\ref{eq: schrod2}) can be combined to give
\begin{equation} \label{eq: complexDiff6}
\frac{\partial}{\partial z_1} 
= \frac{1}{\sqrt{2}} \left[ \frac{\partial}{\partial x_1} - \imath \frac{\partial}{\partial x_2} + \frac{(x_1-\imath x_2)}{\alpha |x_i|} \right],
\end{equation}
\begin{equation} \label{eq: complexDiff7}
\frac{\partial}{\partial z_2} 
= \frac{1}{\sqrt{2}} \left[ \frac{\partial}{\partial x_2} - \imath \frac{\partial}{\partial x_2} + \frac{(x_2-\imath x_1)}{\alpha |x_i|} \right],
\end{equation}
\begin{equation} \label{eq: complexDiff8}
\frac{\partial}{\partial z_3} = \frac{\partial}{\partial x_3} + \frac{x_3}{\alpha |x_i|}
\end{equation}
These results lead to the operator relationship
\begin{equation}\label{eq: qprop1}
\frac{\partial}{\partial x_i^2} = \frac{\partial^* }{\partial z_i} \left( \frac{\partial}{\partial z_i} \right) - \frac{2}{\alpha r}z_i \frac{\partial}{\partial z_i} - \frac{2}{\alpha r} + \frac{1}{\alpha^2} 
\end{equation}
having also used eq. (\ref{eq: complexProd}). It will be convenient to define the $\nabla^2$ operator in complex space to be
\begin{equation} \label{eq: complexNabla}
\nabla^2 = \frac{\partial^* }{\partial z_i} \left( \frac{\partial}{\partial z_i} \right)
\end{equation}
This definition recognises the additional complex terms in eq. (\ref{eq: complexProd}) are uniquely attributable to the use of a complex time and would vanish if $\tau$ were reset to $t$.
 
Eqs. (\ref{eq: complexDiff3}) and (\ref{eq: qprop1}) enable eqs. (\ref{eq: schrod1}) and (\ref{eq: schrod2}) to be rewritten as
\begin{equation}\label{eq: complexSchrod1}
-\frac{\hbar^2}{2m_e}\nabla^2 \Psi + \frac{\hbar}{m_e\alpha r}z_i \frac{\partial \Psi}{\partial z_i} - \frac{1}{r} \left( \frac{e^2}{4 \pi \epsilon} - \frac{\hbar^2}{m_e \alpha} \right)\Psi = 0
\end{equation}
\begin{equation} \label{eq: complexSchrod2}
E\Psi = \imath \frac{\partial \Psi}{\partial \tau} 
\end{equation}
having set
\begin{equation} \label{eq: alphaDef}
\alpha = \sqrt{\frac{-\hbar^2}{2m_eE}} 
\end{equation}
Eqs. (\ref{eq: complexNabla}) and (\ref{eq: alphaDef}) together constitute a description of the hydrogen  atom in terms of $(z_i,\tau)$-coordinates equivalent to eqs. (\ref{eq: schrod1}) and (\ref{eq: schrod2}). It is usual to solve eq. (\ref{eq: schrod1}) using real spherical polar coordinates $(r, \theta, \phi)$. Eq. (\ref{eq: complexSchrod1}) could also be solved in this coordinate system. In this case, we would set
\begin{equation} \label{eq: dotop}
z_i \frac{\partial \Psi}{\partial z_i} = r \frac{\partial \Psi}{\partial r}
\end{equation}
and write $\nabla^2$ in standard $(r, \theta, \phi)$-coordinates. The resulting solution would be equivalent to the solution of eq. (\ref{eq: schrod1}) but expressed in terms of the complex time coordinate $\tau$ instead of the real one $t$.

For the solution of the Schr\"{o}dinger equation (\ref{eq: complexSchrod1}) in this paper it is intended to use complex spherical coordinates $(z_r, z_\theta, z_\phi)$. These have been chosen to take advantage of spherical symmetry but avoid the use of trigonometric functions. These complex spherical polar coordinates are defined to be
\begin{equation}\label{eq: cspc_def}
z_r = |z_i|, \quad z_\theta = \frac{z_3}{|z_i|}, \quad z_\phi = \frac{z_1}{|z_1|} 
\end{equation}
They are related to spherical coordinates in real space through the following expressions:
\begin{equation}\label{eq: cspc2rspc}
r = z_r, \quad \theta = \arccos(z_\theta), \quad \phi = -\imath \ln(z_\phi)
\end{equation}
Eqs. (\ref{eq: cspc2rspc}) are easily checked using eqs. (\ref{eq: conftrans1}) and (\ref{eq: conftrans2}) alongside the relationship $z_1 = |z_1|\exp(\imath\phi)$.

The calculation of the $\nabla^2$-operator in complex spherical polar coordinates is similar to its calculation in real spherical polar coordinates but simpler. It takes the form   
\begin{equation}\label{eq: nabla2_cspc}
\nabla^2 = \frac{1}{z_r^4} \left(z_r^2 \frac{\partial}{\partial z_r} \right)^2 - \frac{\hat{L}^2}{\hbar^2 z_r^2}
\end{equation}
where
\begin{equation}\label{eq: orb_ang_mom_sqr_cspc}
\hat{L}^2 = \frac{-\hbar^2}{(1-z_\theta^2)}\left\{ \left[ (1-z_\theta^2) \frac{\partial}{\partial z_\theta} \right]^2 + \left( z_\phi \frac{\partial}{\partial z_\phi} \right)^2 \right\}
\end{equation} 
is read off as the quantum mechanical orbital angular momentum squared operator. One of the easiest checks on eq. (\ref{eq: nabla2_cspc}) is to transform it in to real spherical polar coordinates using eqs. (\ref{eq: cspc2rspc}).

It will be convenient to assume the Schr\"{o}dinger equation (\ref{eq: complexSchrod1}) has a product solution of the form
\begin{eqnarray} \label{eq: psi_product} 
\Psi(z_r,z_\theta,z_\phi,\tau) = R(z_r)\Theta(z_\theta)\Phi(z_\phi)\exp(-\imath E \tau / \hbar)
\end{eqnarray}
enabling it to be decomposed in to the component equations
\begin{equation}\label{eq: complexSchrod_r}
\xi \frac{d^2 R}{d\xi^2} + (2-\xi)\frac{dR}{d\xi} - \left( 1 - \lambda + \frac{L^2}{\hbar^2 \xi}\right)R = 0
\end{equation}
\begin{equation}\label{eq: complexSchrod_theta}
\frac{d}{dz_\theta} \left[(1-z_\theta^2)\frac{d\Theta}{dz_\theta} \right] + \frac{k^2\Theta}{(1-z_\theta^2)} + \frac{L^2\Theta}{\hbar^2} = 0
\end{equation}
\begin{equation}\label{eq: complexSchrod_phi}
\left( z_\phi \frac{\partial}{\partial z_\phi} \right)^2\Phi - k^2\Phi = 0
\end{equation}
where
\begin{equation}\label{eq: complexSchrod_lambda}
\xi = \frac{2r}{\alpha}, \quad \lambda = \frac{e^2}{4\pi \epsilon \hbar}, \sqrt{\frac{-m}{2E}}
\end{equation}
and $L$ and $k$ are the constants of separation. 

It is notable that $L$ and $k$ must satisfy the eigenvalue equations:
\begin{equation}\label{eq: orb_ang_mom_eq}
\hat{L}^2 \Psi = L^2 \Psi, 
\end{equation}
\begin{equation}\label{eq: z_ang_mom_eq}
\hat{L}_3 \Phi = k \hbar \Phi
\end{equation}
where the form of the z-component of angular momentum operator 
\begin{equation}\label{eq: ang_mom_zform}
\hat{L}_3 = z_\phi \frac{\partial}{\partial z_\phi} 
\end{equation}
can be identified from eq. (\ref{eq: complexSchrod_phi}). 

Eq. (\ref{eq: z_ang_mom_eq}) has the eigensolution $\Phi = z_\phi^k$. This transforms back to $\Phi = \exp(-\imath k\phi)$ in real spherical polar coordinates. Clearly, $k$ must be and integer for $\Phi$ to be a periodic function. 

Eq. (\ref{eq: complexSchrod_theta}) is Legendre's equation \cite{AS} giving $\Theta = P^k_l(z_\theta)$ where $P^k_l$ are associated Legendre polynomials, $L^2=l(l-1)\hbar^2$ and $l (\geq 0)$ is the orbital angular momentum quantum number. In this case, the connection to real spherical polar coordinates is immediate since $z_\theta = \cos(\theta)$.

The solution of eq. (\ref{eq: complexSchrod_r}) is facilitated through the transformation $R = \xi^l R^\prime$ into Laguerre's differential equation. Putting these results together leads to the complete solution of the Schr\"{o}dinger equation for the hydrogen atom in complex $(z_r,z_\theta, z_\phi, \tau)$-coordinates. This takes the form 
\begin{eqnarray} \label{eq: eigensolutions_cspc} 
\Psi_{nlk} =  {\cal{N}}_{nl} \left( \frac{2z_r}{n\alpha}\right)^l L_{n-l-1}^{2l+1}\left(\frac{2z_r}{n\alpha}\right)P^k_l(z_\theta)z_\phi^k\exp(-\imath E\tau / \hbar)
\end{eqnarray}
where ${\cal{N}}_{nl}$ is the normalizing constant and the integer $n (>0)$ is the principal quantum number. The energy eigenvalues for the bound electron
\begin{equation} \label{eq: energy_eigenvalues}
E_n = - \frac{m_ee^4}{32\pi^2 \epsilon^2\hbar^2n^2}
\end{equation}
are determined during the solution of the radial component of the equation.

Eq. (\ref{eq: eigensolutions_cspc}) is readily transformed into $(r, \theta, \phi, \tau)$-coordinates to reveal the form of the hydrogenic wavefunctions most often presented in standard textbooks on quantum mechanics. It is notable, however, that the solution of eqs. (\ref{eq: complexSchrod_r}) through (\ref{eq: complexSchrod_phi}) has taken fewer steps than the solution starting from the initial component form of eq. (\ref{eq: schrod1}) in $(r, \theta, \phi, t)$-coordinates. This is for two reasons. First, the complex time coordinate served a useful purpose in eliminating an exponential term of the form $\exp(-2r/\alpha n)$ in the radial wavefunction. This is otherwise a manual process in  $(r, \theta, \phi, t)$-coordinates in order to reduce the radial problem to Laguerres's equation. Second, the use of complex versus real spherical polar coordinates has stripped away the need to manipulate trigonometric functions.

\section{Concluding Remarks}
The motivation for this paper has been the idea that problems that cannot be represented in one complex plane can be decomposed for representation over multiple complex planes. There are two reasons a problem might require decomposition. One is that it generates fields that do not satisfy the Cauchy-Riemann equations. The other is that the problem is defined in more than two dimensions.

The method that has been found for decomposing problems is to use an isometric transformation to take the problem from a real vector space to a complex one. This term decomposition applies since any point $(x_1, x_2)$ in a plane in the real vector space will transform to $(z_1, z_2)$ and therefore image points in two complex planes. It has been further shown any function $f(x_1, x_2)$ will be holomorphic in both the $z_1$-plane with $z_2$ held constant and the $z_2$-plane with $z_1$ held constant. It is inferred that any problem that can be expressed in a real vector space of $N$-dimensions can also be represented in a complex space composed of no more than $N$ complex planes such that all fields will be holomorphic in the complex planes.

The decomposition principle just outlined has been demonstrated for the case of the Schr\"{o}dinger equation of the hydrogen atom. This well known system has both more than two dimensions and features such as angular momentum that cannot be represented in a single complex plane. It has been nevertheless transformed into a complex space consisting of three complex planes and solved for the complete set of hydrogenic wavefunctions. The solution has utilized complex spherical coordinates that eliminate the need for trigonometric functions. The Laplacian operator has been found to be simpler in complex versus real spherical coordinates facilitating a solution process that overall has many fewer steps in complex space than in real spherical polar coordinates.


\begin{thebibliography}{99}

\bibitem{ZN} Z. Nehari, Conformal Mapping, Dover (1982)

\bibitem{DEB} D.E. Blain, Inversion Theory and Conformal Mapping, American Mathematical Society (2000) 

\bibitem{RJD1} R.J.Ducharme, arXiv:quant-ph/1003.2758

\bibitem{DFL} D.F.Lawden, The Mathematical Principles of Quantum Mechanics, Dover (2005)

\bibitem{RJD2} R.J.Ducharme, arXiv:quant-ph/1003.0717

\bibitem{RJD3} R.J.Ducharme, arXiv:quant-ph/1005.2792

\bibitem{FF} D. Faraoni and D.M. Faraoni, Found. of Phys.,32(5) 2002

\bibitem{AS} M. Abramowitz and I. A. Stegun, Handbook of Mathematical Functions, Dover (1970)

\end{thebibliography}
\end{document}